\documentclass[preprint,aps]{revtex4}
\usepackage{graphics}
\usepackage{graphicx}
\usepackage{bm}
\usepackage{ifpdf}

\begin{document}


\title{Ferroelectric thermal phase transition and polarization precursor dynamics in  Ca$_{x}$Ba$_{1-x}$Nb$_{2}$O$_6$ tungsten bronze type oxides}

\author{Desheng Fu$^{1,2,3}$ and Kensuke Fukusawa$^1$\\
$^1$Department of Electronics and Materials Science,
Faculty of Engineering, Shizuoka University, 3-5-1 Johoku,
Naka-ku, Hamamatsu 432-8561, Japan.\\
$^2$Department of Engineering, Graduate School of Integrated Science
and Technology, Shizuoka University, 3-5-1 Johoku, Hamamatsu
432-8561,
Japan.\\
$^3$Department of
Optoelectronics and Nanostructure Science, Graduate School of
Science and Technology, 3-5-1 Johoku, Naka-ku, Hamamatsu 432-8011,
Japan. \\
E-mail: fu.tokosho@shizuoka.ac.jp}


\begin{abstract} Polycrystals of  Ca$_{x}$Ba$_{1-x}$Nb$_{2}$O$_6$ (CBN) tungsten bronze type  oxides
have been prepared and their structural, dielectric, and thermal
properties have been investigated. It was found that CBN alloys with
ferroelectric tetragonal tungsten bronze structure were only
available in a composition range of $0.19 \leq x \leq 0.32$. It was
also showed that CBN can be classified as a ferroelectric with a
first-order thermal phase transition showing polarization precursor
dynamics before transition into ferroelectric phase, in sharp
contrast to an isostructural alloy Sr$_{x}$Ba$_{1-x}$Nb$_{2}$O$_6$
(SBN) that shows typical relaxor behaviors. The local polarizations
were found to grow exponentially within the paraelectric mother
phase in a large temperature range of $T_{\rm c}<T<T_{\rm c}+
88\sim140$ K on cooling. Furthermore, a phase diagram was
established for CBN ferroelectric alloys. These findings may get an
insight into the true nature of ferroelectric phase transition in
this potential electro-optic material.
\end{abstract}


\maketitle

\section{Introduciton}

There are increasing interests in seeking  novel electro-optic
materials analogous to the famous Sr$_{x}$Ba$_{1-x}$Nb$_{2}$O$_6$
(SBN, $0.32\leq x\leq 0.75$) that  belongs to the
(A$_1)_2$(A$_2)_4$(B$_1)_2$(B$_2)_8$O$_{30}$ tetragonal tungsten
bronze (TTB) structure  with the non-centrosymmetry of P4{\it bm} at
room temperature (see  Fig.\protect\ref{Fig0})
\protect\cite{Podlozhenov,Chernaya1,Chernaya2} and shows excellent
electro-optic and pyroelectric properties \protect\cite{Glass} and typical
relaxor behaviors.\protect\cite{Kleemann1,Kleemann2} Recently,
Ca$_{x}$Ba$_{1-x}$Nb$_{2}$O$_6$ (CBN, $0.2\leq x\leq 0.4$) single
crystals have received considerable critical attention due to their
excellent electro-optic properties and  higher working temperature
in comparison with SBN.\protect\cite{Ebser,Burianek,Muehlberg,Qi,Song} The
ferroelectricity of CBN is first discovered  by Smolenskii et
al..\protect\cite{Smolenskii} CBN alloys also have  the ferroelectric TTB
structure at room temperature, however,  contrary to Sr in SBN where
Sr ions randomly occupy both the two large A$_1$ and A$_2$ sites in
the tungsten bronze framework of NbO$_6$ octahedra, smaller Ca ions
in CBN only occupy the A$_1$ sites in the structure. It has been
demonstrated experimentally that smaller A$_1$ site is almost
exclusively occupied by Ca while  larger A$_2$ site is predominately
by Ba in CBN single crystal with congruent melting composition
($x$=0.28).\protect\cite{Graetsch1,Graetsch2} In ferroelctric CBN alloys,
niobium atoms are displaced from the centers of their coordination
polyhedra and shifted along the tetragonal $c$-axis, which is
considered to be the origin of  spontaneous electric
polarization.\protect\cite{Chernaya2, Abrahams}

Currently, most of investigations on CBN alloys are focused on the
crystal growth\protect\cite{Ebser,Song}, dielectric,\protect\cite{Qi,Niemer}
ferroelectric,\protect\cite{Qi} optic,\protect\cite{Ebser2,Heine,Sheng} and elastic
\protect\cite{Pandey,Pandey2,Pandey3,Suzuki} properties  of CBN single
crystal with congruent melting composition $x$=0.28. Despite these
investigations on CBN, there is still a lack of full understanding
on the basic issues in CBN alloys. One important problem is that CBN
alloys are   relaxors (such as their isostructural compounds SBN
alloys\protect\cite{Kleemann2,Blinc,Banys,Miga}), which is characterized by
a broad peak of dielectric susceptibility with strong frequency
dispersion in the radio frequency over a large temperature range
\protect\cite{Miga,Levstik,Fu1,Fu2} and  smear ferroelectric phase
transition\protect\cite{Fu1,Taniguchi} without an obvious heat capacity
peak,\protect\cite{Moriya} or they are  ferroelectrics (such as the normal
ferroelectric BaTiO$_3$), which have a well-defined
para-ferroelectric  thermal phase transition at Curie point $T_{\rm
c}$ \protect\cite{Fu3} but show polarization precursor dynamics before
transition into the ferroelectric
phase.\protect\cite{Burns,Tai,Ziebinska,Ko,Dulkin,Pugachev} Since CBN is an
isostructural compound of SBN that typical relaxor phenomena have
been clearly demonstrated,\protect\cite{Banys,Miga} it is naturally expected
that CBN should show the relaxor behaviors such as a broad peak of
dielectric susceptibility with strong frequency dispersion  due to
the  dynamics of polar-nano-regions (PNRs) occurring in the
paraelectric mother phase. Following this scenario, several attempts
have been made to confirm the relaxor behaviors in
CBN.\protect\cite{Pandey,Pandey2,Pandey3,Suzuki} From the investigations on
lattice strain and thermal expansion of CBN single crystal with
congruent melting composition $x$=0.28, Pandey et al. found the
deviation from the linear temperature dependence of lattice strain
and the anomalous thermal expansion in this single crystal.\protect\cite{Pandey} They
attribute these anomalous elastic behaviors to the relaxor phenomena
occurring in CBN crystal. They further suggest  Burns
temperature \protect\cite{Burns} $T_{\rm B}$ to be 1100 K, which
characterizes the initiation of dynamic PNRs, and the intermediate
temperature $T^*$ to  be 800 K,  which indicates the beginning of
PNRs freezing,\protect\cite{Roth,Dkhil}  for CBN with $x=0.28$.\protect\cite{Pandey}
On the other hand, in  a recent Brillouin scattering study, $T_{\rm
B}$ and $T^*$ are proposed to be approximately 790 K and 640 K,
respectively, for CBN with the same composition.\protect\cite{Suzuki} The
difference in     $T_{\rm B}$ and $T^*$ values estimated by
different techniques for a same compound is surprisingly large.
Apparently, the existence and  exact values of $T_{\rm B}$ and $T^*$
in CBN need to be clarified in further investigations.

It should be noticed that the existence of $T_{\rm B}$ and $T^*$ is
not the characteristic properties of relaxor. In a normal
ferroelectric such as BaTiO$_3$, the existence of these two
characteristic temperatures have been clearly
demonstrated.\protect\cite{Burns,Dulkin} In
 early investigations, Burns et al.\protect\cite{Burns} have demonstrated that
the temperature dependence of the optic index of refraction, $n(T)$,
deviates from the high temperature extrapolated value between
$T_{\rm c}$ and $T_{\rm c} + 180$ K in the paraelectric phase of
BaTiO$_3$ crystal. This deviation is commonly accepted to be due to
the local polarization precursor dynamics  before the
para-ferroelectric phase transition in BaTiO$_3$, thus the
temperature where this deviation occurs is generally called as Burns
temperature $T_{\rm B}$, which is widely used to characterize the
appearance of PNRs in the paraelectric phase of ferroelectric or
relaxor. In additional to the optical
measurements,\protect\cite{Burns,Ziebinska,Pugachev} various other
experiments such as pulsed X-Ray laser
measurements,\protect\cite{Tai,Namikawa} acoustic emission\protect\cite{Dulkin} and
Brillouin light scattering\protect\cite{Ko} all show the existence of local
polarization in BaTiO$_3$ before transition into the ferroelectric
phase. Theoretically, it has been proposed that local dynamical
precursor domains are common to perovskite ferroelectrics, and these
polar domains grow upon approaching $T_{\rm c}$ to coalesce into a
homogeneously polarized state at $T_{\rm
c}$.\protect\cite{Bussmann-Holder1,Bussmann-Holder2} Similarly, PNRs develop
from $T_{\rm B}$ in relaxor, and cause the very strong frequency
dispersion of dielectric susceptibility. In contrast to relaxor,
precursor domains in the perovskite ferroelectric such as BaTiO$_3$
do not need to give rise to a frequency-dependent dielectric
response in the radio frequency.\protect\cite{Bussmann-Holder1}

Furthermore, a sharp peak of heat capacity has been observed at the
para-ferroelectric phase transition in CBN  crystal with
$x=0.311$.\protect\cite{Muehlberg} Also, in the dielectric measurements on
CBN single cyrstal with congruent melting composition $x$=0.28, it
seems that the para-ferroelectric phase transition temperature is
not dependent on  frequency.\protect\cite{Qi} All these findings in CBN  are
in sharp contrast to the characteristic properties of relaxor such
as  smear phase transition \protect\cite{Fu1,Taniguchi} without an evident heat
capacity peak,\protect\cite{Moriya} and very strong frequency dependence
of dielectric responses in the radio frequency around a temperature,
$T_{\rm m}$, where the maximum of dielectric response occurs. These
results suggest that CBN alloys may be classified as a
ferroelectric with thermal phase transition associating with
polarization precursor dynamics rather than a relaxor.

In this study, we firstly determined the solid solution limit of CBN
ferroelectric alloys. We then investigated the para-ferroelectric
phase transition in these  alloys by the dielectric measurements and
the differential scanning calorimetry (DSC). Contrary to the relaxor
picture what have been expected in previous
investigations,\protect\cite{Pandey,Pandey2,Pandey3} we clearly demonstrated that CBN
alloys can be classified as ferrolectrics with a first-order thermal
phase transition associating with precursor dynamics over a large
temperature range above $T_{\rm c}$ such as
BaTiO$_3$.\protect\cite{Burns,Ziebinska,Pugachev} We also found that the
local polarizations grown exponentially for a temperature range of
$T_{\rm c}<T<T_{\rm c}+ 88\sim140$ K on cooling in CBN alloys. These findings are
of benefit in understanding the nature of ferroelectric phase
transition in  CBN crystals that are currently developed for the
electro-optic applications.

\section{Experimental}

We used a conventional solid-state reaction method to prepare  CBN
polycrystals, which allows us to easily control composition of CBN
samples. To obtain dense ceramics for electrical measurements, CBN ceramics samples were fired
3 hours at a temperature of 1693 K (approximately 50 K lower than
its melting point\protect\cite{Muehlberg}). This firing allows us to obtain
ceramics samples with a relative density larger than 93 \% of the
density calculated by the structural data. Grain size in the
ceramics samples was observed to be  approximately 10$\mu$m $\sim$ $50\mu$m by Scanning Electron Microscopy. To determine the solid
solution limit, ceramics samples with composition of $0.15\leq
x\leq0.40$ were prepared and investigated. Phase formation and
lattice parameters of CBN were determined by powder X-ray
diffraction technique using the Cu-K$\alpha$ radiation
($\lambda=1.5418$ {\AA}). Ceramics samples coated with gold electrode
were used for the dielectric measurements by using Agilent 4980A LCR
meter. The dielectric measurements were carried out in a
temperature range of 300 K - 870 K with a heating/cooling rate of 2
K/min for the exact determination of   ferroelectric transition
temperature in CBN alloys. To confirm the thermal phase transition
in CBN, we also performed DSC measurements by using SII DSC6220
differential scanning calorimetry.

\section{Results and Discussions}

\subsection{Ferroelectric solid solution limit and  lattice parameters}

To determine the  solid solution limit of CBN ferroelectric alloys,
we have prepared the samples with  composition ranging from $x=0.15$ to
$x=0.40$ at a firing temperature of 1693 K, and examined the phase
formation by powder x-ray diffractions. As shown in
Fig.\protect\ref{fig1}, single phase of CBN with ferroelectric TTB structure
is  only available in a composition range of $0.19\leq x\leq0.32$, beyond
which CaNb$_2$O$_6$- or BaNb$_2$O$_6$-type phase with
non-ferroelectric orthorhombic structure was  found in the
compounds. The change of lattice parameters or Curie temperature
with composition also show a saturation for $x<0.19$ or $x>0.32$.
These facts indicate that single phase of CBN ferroelectric alloys
is available only in a composition range of $0.19\leq x\leq0.32$.

The lattice parameters of CBN  alloys at room temperature were then
determined by the method of least squares using  twelve reflections
with 2$\theta>50$ degree. The results are shown in Fig.\protect\ref{fig2}. In
CBN  alloys with  ferroelectric TTB structure, $a$-axis lattice
is nearly unchanged with composition within the error range. In contrast, the polar
$c$-axis lattice is shorten with an increase in Ca concentration,
which results in a similar decrease of unit cell volume. This result
is expectable because the ionic radii of Ca and Ba are 1.34 {\AA}
and 1.61 {\AA},\protect\cite{Shannon} respectively, and Ca has a smaller
ionic radius. In spite of the large change in composition $x$ ($\Delta
x=0.13$), the corresponding change in the unit cell volume is
surprisingly small (approximately 0.65\%). This situation is
different significantly from the case of substitution of Ca for Ba
in Ba$_{1-x}$Ca$_x$TiO$_3$ perovskite oxides, which shows an
approximate 3.5\%-reduction in the unit cell volume at a same level
of Ca-substitution.\protect\cite{Fu3} This fact seems to suggest that the
stacking structure in TTB-type CBN crystal is mainly determined by
the framework of NbO$_6$ octahedra as shown in Fig.(\protect\ref{Fig0}).

\subsection{Thermal phase transition}

To investigate the para-ferroelectric phase transition in CBN
alloys, we observed the variation of dielectric susceptibility as a
function of temperature in a frequency range from 100 Hz to 100 kHz.
The results were shown in Fig.\protect\ref{fig3} and Fig.\protect\ref{fig4},
respectively. The dielectric loss for lower frequencies ($\leq$ 1 kHz) was
large for temperatures higher than approximately 500 K and was not shown here,  which is
very likely due to the thermal activation of vacancies in the
samples. The large dielectric loss made difficult to determine a
reliable value of dielectric susceptibility for these low frequencies for
the high temperatures. However, as shown in Fig.\protect\ref{fig4}, we
indeed observed that (1) Curies point is not dependent on frequency  ranging from 100 Hz to 100 kHz,
and (2) the dielectric susceptibility measured at frequency between
10 KHz and 100 kHz is nearly independent of frequency in the
temperature range of 300 K - 870 K. We thus consider that the
dielectric susceptibility measured at frequency between 10 KHz and 100 kHz is free from
the effects of vacancies and reflects the intrinsic dielectric
response of  CBN alloys, which will be used in the later analysis.
Apparently,  the dielectric response with frequency and
temperature in CBN alloys  is in sharp contrast to  that observed in
relaxor.\protect\cite{Fu1} Instead, the dielectric behaviors in CBN are
similar to those observed in  normal ferroelectric such as
BaTiO$_3$.\protect\cite{Fu3}

The temperature dependence of dielectric susceptibility shown in
Fig.\protect\ref{fig4} clearly indicates that the para-ferroelectric phase
transition has a thermal hysteresis.  Curies points on heating and
cooling  have different values. The difference of $T_{\rm c}$
between heating and cooling is 12.4 K for $x=0.19$ and it increases
to 25.2 K for $x=0.32$. This indicates that the substitution of Ba
with Ca in CBN enhances the thermal hysteresis of para-ferroelectric
phase transition. On the other hand, this substitution lowers
Curies point $T_{\rm c}$ of CBN alloys as demonstrated clearly in
Fig.\protect\ref{fig3}.

To further confirm the nature of thermal phase transition in CBN
ferroelectric alloys, we also observed the change in enthalpy during
the phase transition by the DSC measurements. The results are shown
in Fig.\protect\ref{fig5} Although our equipment doesn't allow us to
determine the heat capacity of phase transition, we really
observed a change in enthalpy during the phase transition. DSC
measurements also clearly  indicate that the phase transition has a thermal
hysteresis. This result is in good agreements with that observed in
the dielectric measurements shown in Fig.\protect\ref{fig3}. Our result is
also well accorded with that reported for CBN crystal with $x=0.31$
 showing a sharp peak of heat capacity during the phase
transition.\protect\cite{Muehlberg}

All facts showed above indicate that CBN alloys undergo a thermal
phase transition. It can be concluded that the para-ferroelectric
phase transition in CBN alloys is of first-order and has a thermal
hysteresis of 12.4 K to 25.2 K depending on  Ca-concentration.

\subsection{Precursor behaviors}

In a recent investigation on CBN single crystal with congruent
melting composition $x$=0.28, the birefringence has been
demonstrated to occur in a temperature region of  $T_{\rm
c}<T<T_{\rm c}+ 40 $K.\protect\cite{Ebser2} The birefringence of CBN with
TTB structure might have arisen due to the anisotropy of
polarization fluctuation $\langle P_c^2\rangle- \langle
P_a^2\rangle$. In  this case,  the birefringence
reads\protect\cite{Lehnen,Takagi}
 \begin{equation}\label{eq1}
\Delta n=-(n^3_0/2)( g_{11}-g_{12})\langle P^2 \rangle,
 \end{equation}
where $g_{11}$ and  $g_{12}$  are the electro-optic coefficients.
The occurrence of birefringence provides a direct evidence to the
existence of local polarization in the paraelectric phase of CBN
crystal.

To get further insights on the polarization precursor dynamics
occurring for $T>T_{\rm c}$, we performed an analysis on the
temperature dependence of dielectric susceptibilities of CBN alloys
within its solid solution limit. As mentioned in the  above section,
here, we only used the data obtained at 100 kHz since these data are
considered to be free from the effects of vacancies on the
dielectric response. As shown in Fig.\protect\ref{fig6}(b), the dielectric
susceptibility $\chi'$ obeys Curie law,
\begin{equation}\label{eq2}
    \chi'=\pm C/(T-T_0)\;\; {\rm for}\;\; T<T_{\rm c}\;\;  {\rm or}\;\; T>T_{\rm
    B},
\end{equation}
where $C$ is Curie constant and $T_0$ is Curie-Weiss temperature.
Upon cooling, the  dielectric susceptibility deviates from  Curie
law at a characteristic temperature. Since the existence of Burns
temperature $T_{\rm B}$ and  the intermediate temperature $T^*$ are
not well established for CBN,\protect\cite{Pandey,Suzuki} Here, we
tentatively consider this characteristic temperature as Burns
temperature $T_{\rm B}$. The values of $T_{\rm B}$ were estimated to
be approximately $T_{\rm c}$+88 K for $x=0.19$ and $T_{\rm c}$+143 K
for $x=0.32$, respectively, and $T_{\rm B}$ showed a nearly linear change
with composition as shown
in the phase diagram (Fig.\protect\ref{fig9}(a)). The fitting parameters
obtained from Curie law are summarized in Fig.\protect\ref{fig7}. It should
be noticed that Curie constant in the paralectric phase ($T>T_{\rm
B}$) of CBN alloys has a similar magnitude of that of BaTiO$_3$
single crystal ($1.5 \times10^5$ K).\protect\cite{Shiozaki}

For temperatures higher than $T_{\rm B}$, the dielectric
susceptibility of  CBN alloys exactly follows Curie law, which
indicates that the dielectric response in these high temperatures
can be considered to be due to the lattice dynamics. However, on
cooling, deviation from Curie law  was observed from $T\approx
T_{\rm B}$. As mentioned above, this deviation from Curie law of the
dielectric susceptibility can be reasonably attributed to the
polarization precursors dynamics occurring before para-ferroelectric
phase transition in CBN alloys. Therefore, it can be considered that
the total dielectric susceptibility is contributed by  both lattice
and precursor dynamics for $T_{\rm c}<T<T_{\rm B}$. The contribution
of lattice dynamics to dielectric responses can be estimated from
Curie law, here, we denote this part of dielectric susceptibility as
$\chi'_{\rm Curie}$. The contribution of  precursor dynamics to the
dielectric response was then estimated by subtracting $\chi'_{\rm
Curie}$ from the total dielectric susceptibility $\chi'$, and has
the form of $\Delta\chi'=\chi'-\chi'_{\rm Curie}$. The temperature
variation of $\Delta\chi'$ is shown in Fig.\protect\ref{fig6}(a) and (c) for
the two end compositions of CBN alloys. Upon cooling, $\Delta\chi'$
increases rapidly, for example, for $x=0.32$, it increases from
nearly zero at $T_{\rm B}$ to a large value that is about 34 \% of
the total dielectric response (2240) at $T_{\rm c}$. For another end
composition $x=0.19$ , precursor dynamics  contributes approximately
25\% of the total dielectric response at $T_{\rm c}$. As shown in
Fig.\protect\ref{fig8}(a), Ca-substitution  enhances the contribution of
precursor dynamics to the total dielectric response in CBN alloys.

Surprisingly, it was found that $\Delta\chi'$ increases
exponentially upon cooling from $T_{\rm B}$ to $T^+_{\rm c}$. This
can be clearly seen in Fig.\protect\ref{fig6}(c) where the {\it y}-axis was
plotted in a logarithmic scale. The data of CBN alloys within the
solid solution limit can be well analyzed by the following
exponential law,
 \begin{equation}\label{eq3}
    \Delta\chi'=\Delta\chi'_{T_{\rm c}}e^{-k_B(T-T_{\rm c})/E_a}
     \end{equation}
where $k_B$ is Boltzmann constant, $\Delta\chi'_{T_{\rm c}}$ and
$E_a$ are constants. It should  be noticed that the exponential law of  dielectric response
with temperature also has been observed in Pb-based ferroelectric
oxides in our recent investigation.\protect\cite{Zhao} The obtained
parameters for equation (\protect\ref{eq3}) were shown in Fig.\protect\ref{fig8}(a)
and (b).

From textbook, we know the relation of polarization $P$ and
dielectric susceptibility $\chi'$,
 \begin{equation}\label{eq4}
    P=\chi'\epsilon_0E,
     \end{equation}
where $E$ is an electric field and $\epsilon_0$ is the dielectric
permittivity  of vacuum. With the combination of this relationship and
the exponential law  shown in equation (\protect\ref{eq3}), it can be
predicted that  local polarizations grow exponentially upon cooling
from $T_{\rm B}$ to $T^+_{\rm c}$ in CBN alloys. Here, $E_a$ is
considered to be thermal activation energy required for the  local
polarization growth before  transition into the ferroelectric phase.
$E_a$ was evaluated to be 2.2 mev $\sim$ 3 mev for CBN alloys, which
corresponds to a temperature of 25 K $\sim$ 30 K. This barrier
required for the precursor growth seems to be reasonable although
there are no available  data in the literatures to our
knowledge.

We further used equation (\protect\ref{eq4}) to make an estimation of local polarization growing at $T=T_{\rm c}^+$ using the very
weak electric field ($E=10^{-3}$ kV/cm) used in the dielectric
measurements. The estimated values were given in Fig.\protect\ref{fig8}(c)
as a function of composition $x$. At $T=T_{\rm c}^+$, the local
polarization is predicted to grow to a magnitude of 0.03 nC/cm$^2$ $\sim$
0.07 nC/cm$^2$ in CBN ceramics.  Ca-substitution seems to enhance
the local polarization growth in CBN alloys. This is reasonable
since Ca-substitution reduces  Cuire point to a lower temperature,
resulting in a larger temperature range ($ T_{\rm c}<T<T_{\rm B}$)
for the local polarization growth as shown in Fig.\protect\ref{fig6}(c) and
Fig.\protect\ref{fig9}(a).

\subsection{Phase diagram}
 Having discussed the thermal phase transition and the precursor
 behaviors in the above sections, we can establish a phase diagram
 for CBN alloys now. The proposed phase diagram was shown in
 Fig.\protect\ref{fig9}(a), in which the transition temperature $T_{\rm c}$ on cooling
 or heating was determined from the temperature dependence of
 dielectric susceptibility and $T_{\rm B}$ was estimated from
 deviation of Curie law in the dielectric susceptibility obtained
 upon cooling.

 In our phase diagram,  CBN alloys can be considered as
 paralectrics with a centrocymmetric tetragonal structure as
 that of isostructural compounds SBN for $T>T_{\rm B}$. On cooling to $T\approx T_{\rm
 B}$, polarization precursors emerge in the paraelectric mother
 phase of CBN alloys. On further cooling, these local
 polarizations growth exponentially with the temperature before
 transition into a ferrolectric phase with a non-centrocymmetric tetragonal  structure  at
 $T=T_{\rm c}$. This ferroelectric thermal phase transition is of
 first-order, and the transition temperature upon heating is
 $12.4\sim 25.2$ K higher than that on cooling. As shown in the
 phase diagram, increase in Ca-concentration leads to a larger thermal
 hysteresis together with the lowering of $T_{\rm c}$ and $T_{B}$. On the
 other hand, the temperature range of precursor existence becomes
 larger as increasing the amount of substitution  of  smaller Ca ions for Ba
 ions. It is still unclear why Ca-substitution enlarges the precursor growth region, the
 local polarization developed at $T_{\rm c}^+$, and  the thermal
 hysteresis of phase transition in CBN alloys.
 A likely reason is relative to the increase of vacancies in
 larger A$_2$ site in the TTB structure due to the increase of  substitution amount of
 Ca for Ba. Because Ca almost exclusively occupied smaller A$_1$
site in CBN TTB structure, increase of Ca-concentration naturally
results in the increase of vacancies in
 A$_2$ site occupied initially by Ba.\protect\cite{Graetsch1,Graetsch2} Increase of vacancies in larger A$_2$ site
with $x$ may give rise to larger fluctuation of local lattice
distortion in CBN alloys, which is though to  be the sources of
local polarizations. This suggestion is supported by  recent
investigations on the variation of lattice strain with $x$ in CBN
alloys, in which the thermal expansion along the polar axis becomes
larger as increasing  Ca-concentration.\protect\cite{Pandey3}

On the other hand, substitution of Ca for Ba leads to a nearly
linear reduction of $T_{\rm c}$ and $T_{B}$ with $x$. As well known
in many normal ferroelectrics, pressure can significantly reduce the
ferroelectric transition temperature. In solid solution, pressure
can be due to chemical substitution of smaller ions for  original
larger one, and this kind of pressure is normally called as chemical
pressure. Chemical-pressure effects on the reduction of $T_{\rm
c}$ have been well studied in many ferroelectric alloys with
perovskite structure such as
(Ba$_{1-x}$Sr$_x$)TiO$_3$,\protect\cite{Shiozaki} in which  $T_{\rm
c}$-reduction is generally considered to be due to the reduction of
unit cell volume by the chemical pressure effects.  As mentioned in
the above section, in CBN alloys, substitution of smaller Ca ions
for Ba ions only lead to an extremely small change  in its unit cell
volume ($\Delta=(V(x)-V(0.19))/V(0.19)$), and  $\Delta$ is less than
0.65\% within the solid solution limit. Thus, it can be considered
that the chemical pressure effects on the reduction of $T_{\rm c}$
was vanishingly small in CBN alloys.

The spontaneous polarization in CBN has been reported to be
originated from  Nb-displacement along the $c$-axis in the TTB
structure.\protect\cite{Graetsch1} It can is expected that the lattice
distortion along the $c$-axis has influence on  Nb-displacement.
Figure\protect\ref{fig9}(b) showed the variation of $T_{\rm c}$ with
$c$-axis lattice constant. Indeed, one can see that $T_{\rm c}$ is
reduced with the shrink of  $c$-axis lattice. This fact indicates
that  the shrink of  $c$-axis lattice should reduce the space
available for  Nb-displacement, thus leading to the reduction of
ferroelectricity and its $T_{\rm c}$. However, since the reduction
in  $c$-axis lattice constant is approximately 0.03 ${\AA}$
within the solid solution, this lattice shrink is not enough to
completely explain the large reduction in $T_{\rm c}$ in CBN alloys.
It seems that other reasons may be existed for larger reduction of
$T_{\rm c}$ with Ca-concentration in CBN alloys. The exact reason for this large reduction in $T_{\rm c}$ with Ca-substitution in CBN alloys remains to be
clarified in  further structural investigations.

\section{Summary}

In summary, we have studied the phase transition in CBN alloys
within its solid solution limit ranging from $x=0.19$ to $x=0.32$.
In contrast to their isostructural compounds SBN that  normally show
typical relaxor behaviors, CBN alloys exhibit  behaviors such as
normal ferroelectric BaTiO$_3$, but show precursors dynamics before
transition into the ferroelectric phase. For $T>T_{\rm B}$, CBN
alloys obey classical Curie law and can be considered to be
paraelectric. On further cooling toward $T_{\rm c}$, local
polarizations occurs in the paraelectric mother phase, and these
polarization precursors grow exponentially as temperature lowering.
Finally, a ferroelectric phase transition was realized at $T_{\rm c}$. This thermal
ferroelectric phase transition is essentially of first-order. A
phase diagram was then establish for CBN ferroelectric alloys. These
findings  provide new insights on understanding the underlying
physics in CBN ferroelectric alloys.



\newpage 


\newpage
\begin{figure}
\caption{\label{Fig0}
(A$_1)_2$(A$_2)_4$(B$_1)_2$(B$_2)_8$O$_{30}$  TTB  structure of
Ca$_x$Ba$_{1-x}$Nb$_2$O$_6$ in projection along the polar $c$-axis,
where $A_1$, $A_2$ and $B_1$/$B_2$ sites are occupied by Ca, Ba and
Ti ions , respectively.}
\end{figure}

\newpage
\begin{figure}
\caption{\label{fig1}  Powder X-ray diffraction patterns of
Ca$_x$Ba$_{1-x}$Nb$_2$O$_6$, which indicate that the ferroelectric
solid solution has a limit of solid solution of $0.19\ll x \ll 0.32$
beyond which single phase is not available.}
\end{figure}

\newpage
\begin{figure}
\caption{\label{fig2} Change of lattice parameters of
Ca$_x$Ba$_{1-x}$Nb$_2$O$_6$ with composition.}
\end{figure}

\newpage
\begin{figure}
\caption{\label{fig3} The dielectric susceptibility of
Ca$_x$Ba$_{1-x}$Nb$_2$O$_6$ polycrystals as a function of
temperature and composition.}
\end{figure}

\newpage
\begin{figure}
\caption{\label{fig4} Examples of  frequency dependence of
dielectric susceptibility and loss for  Ca$_x$Ba$_{1-x}$Nb$_2$O$_6$
polycrystals. It is clear that the transition temperature is not
dependent on frequency and has a thermal hysteresis.}
\end{figure}

\newpage
\begin{figure}
\caption{\label{fig5}  Thermal phase transition in
Ca$_x$Ba$_{1-x}$Nb$_2$O$_6$ detected by differential scanning
calorimetry (DSC). Exotherm or endotherm was seen during the phase
transition as indicated by the arrows. }
\end{figure}

\newpage
\begin{figure}
\caption{\label{fig6}  (a) Examples of  temperature variations of
the measured dielectric susceptibility $\chi'$, the fitting results
by Curie law $\chi'_{\rm Curie}$ (dashed lines), and their
difference $\Delta\chi'=\chi'-\chi'_{\rm Curie}$. (b) Inverse of the
dielectric susceptibility $\chi'$. The fitting results  by Curie law
are also shown by the dashed lines. Deviation from Curie law was
seen from $T\approx T_{\rm B}$ upon cooling. (c) Change of $\Delta
\chi'=\chi'-\chi'_{\rm Curie}$ plotted in a logarithmic scale for
$T_{\rm c}<T<T_{\rm B}$. Dotted lines are the fitting results of an
exponential law given in equation (\protect\ref{eq3}).}
\end{figure}

\newpage
\begin{figure}
\caption{\label{fig7}  Composition change of the fitting  parameters
used in Curie law (equation (\protect\ref{eq2})) for the ferroelectric phase
($T<T_{\rm c}$) and the paraelectric phase for $T>T_{\rm B}$. }
\end{figure}

\newpage
\begin{figure}
\caption{\label{fig8} (a) Composition change of $\Delta
\chi'=\chi'-\chi'_{\rm Curie}$  and $\Delta \chi'/\chi'$  at
$T=T_{\rm c}^+$. (b) Evaluated activation energy required for  the
local  polarization precursors to grow in $T_{\rm c}<T<T_{\rm B}$ in
the paraelectric mother phase. (c) Polarization growing at $T=T_{\rm
c}^+$, which was estimated  by using equation (\protect\ref{eq4}) with the
very weak  field ($E=10^{-3}$ kV/cm) used in the dielectric measurements. }
\end{figure}

\newpage
\begin{figure}
\caption{\label{fig9} (a) Phase diagram proposed for
Ca$_x$Ba$_{1-x}$Nb$_2$O$_6$ ferroelectric alloys. In the shaded
area, local polarizations were expected to grow exponentially with
equation (\protect\ref{eq3}) within the paraelectric mother phase.
FE=ferroelectric and PE=paraelectric. (b) Change of phase transition
temperature with the polar $c$-axis lattice constant. }
\end{figure}

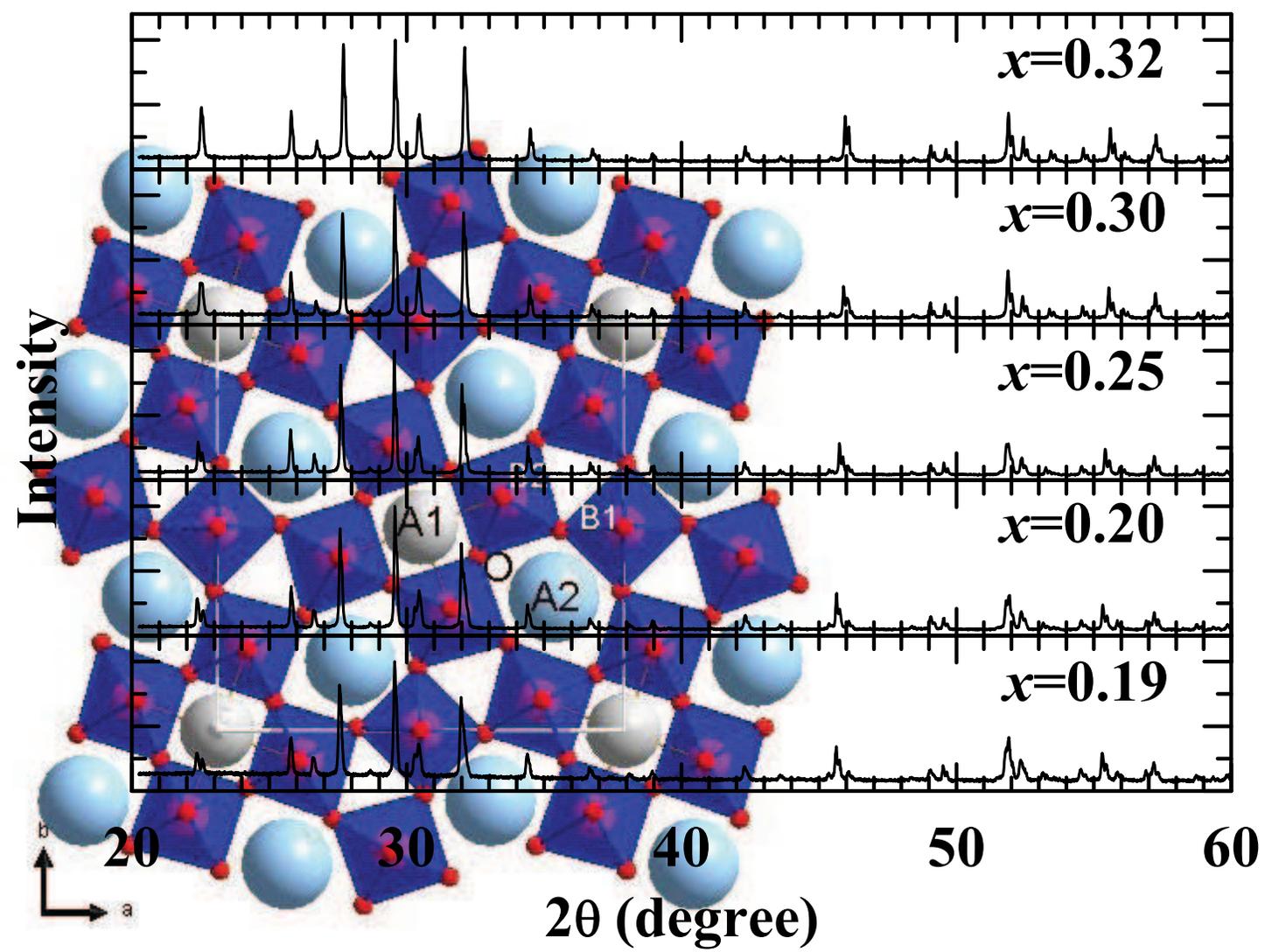
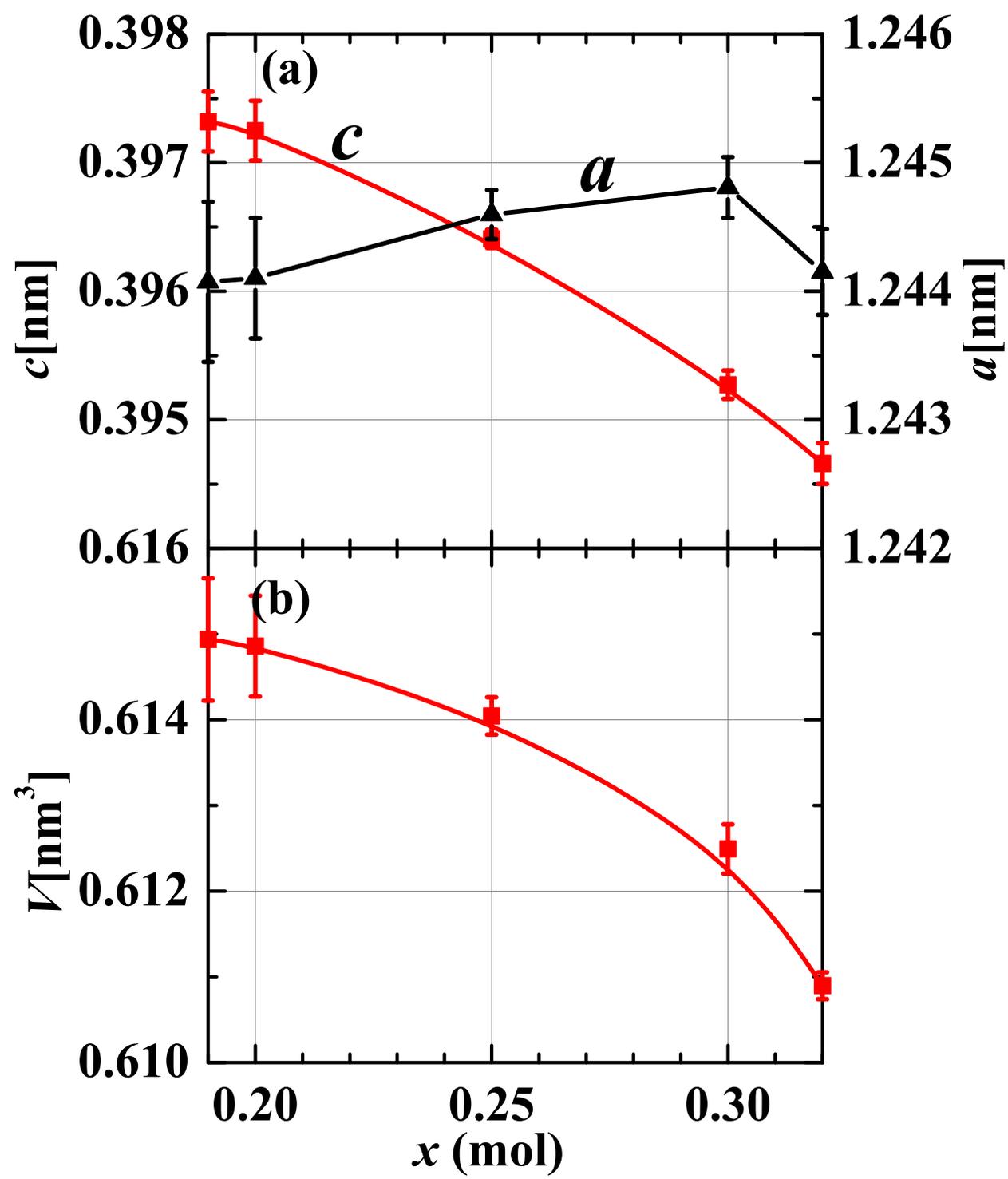
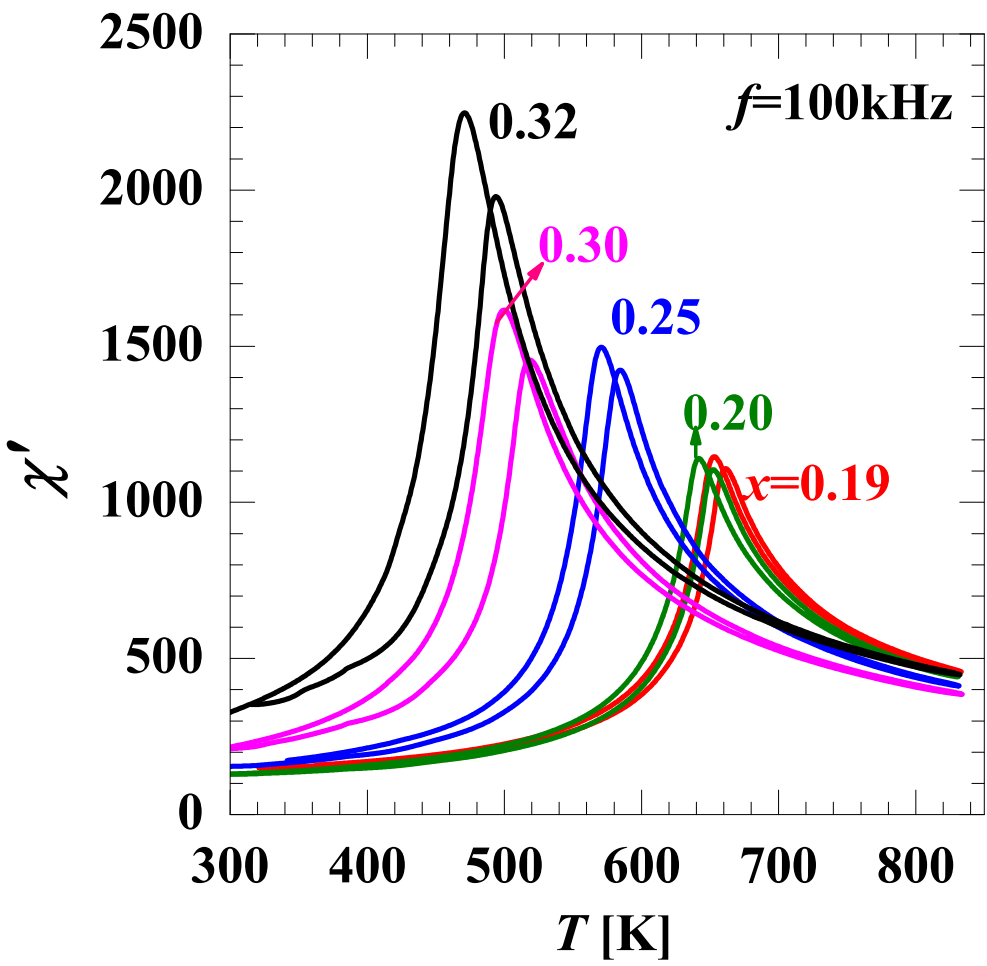
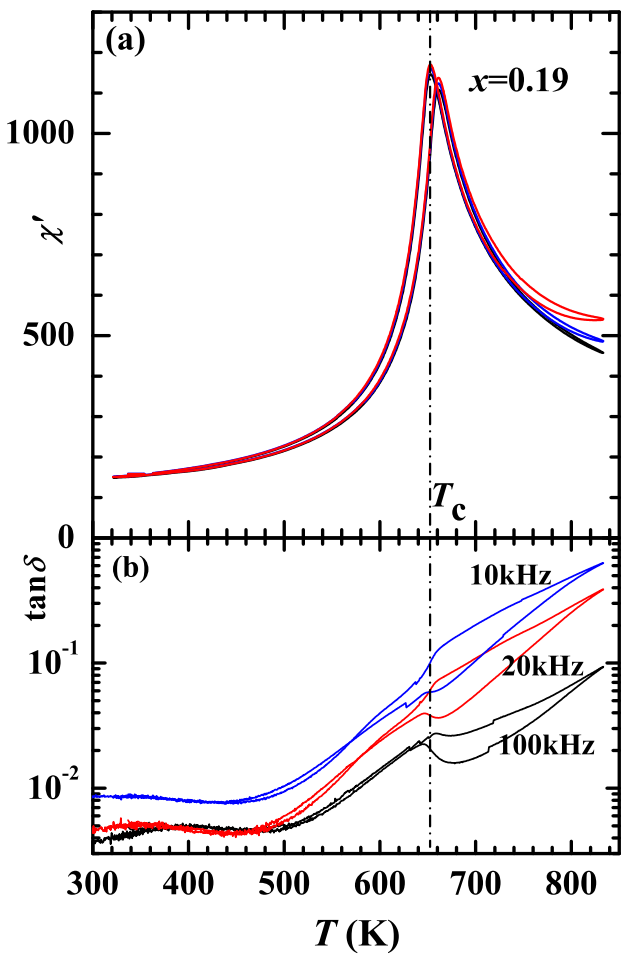
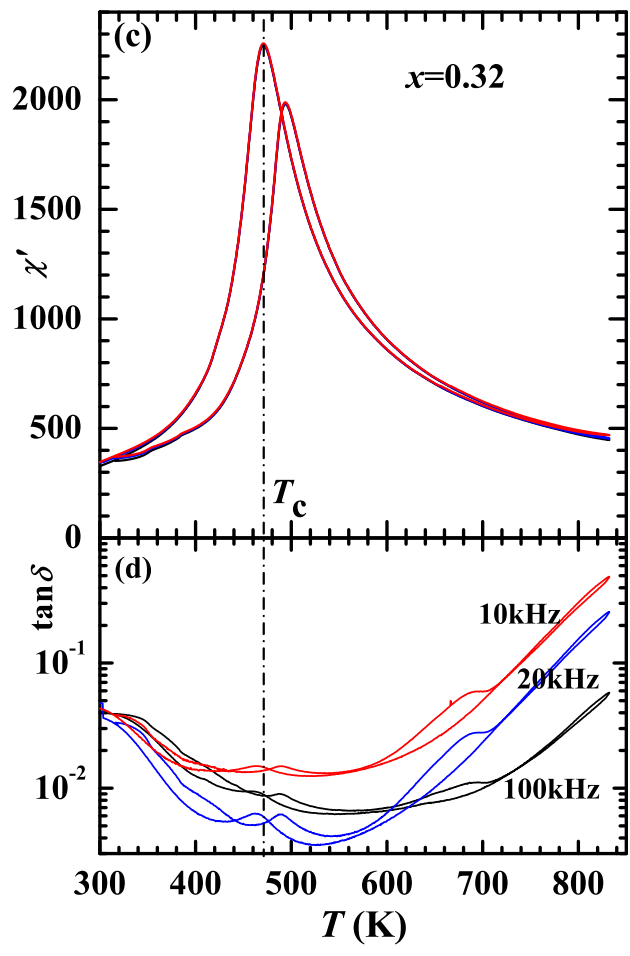
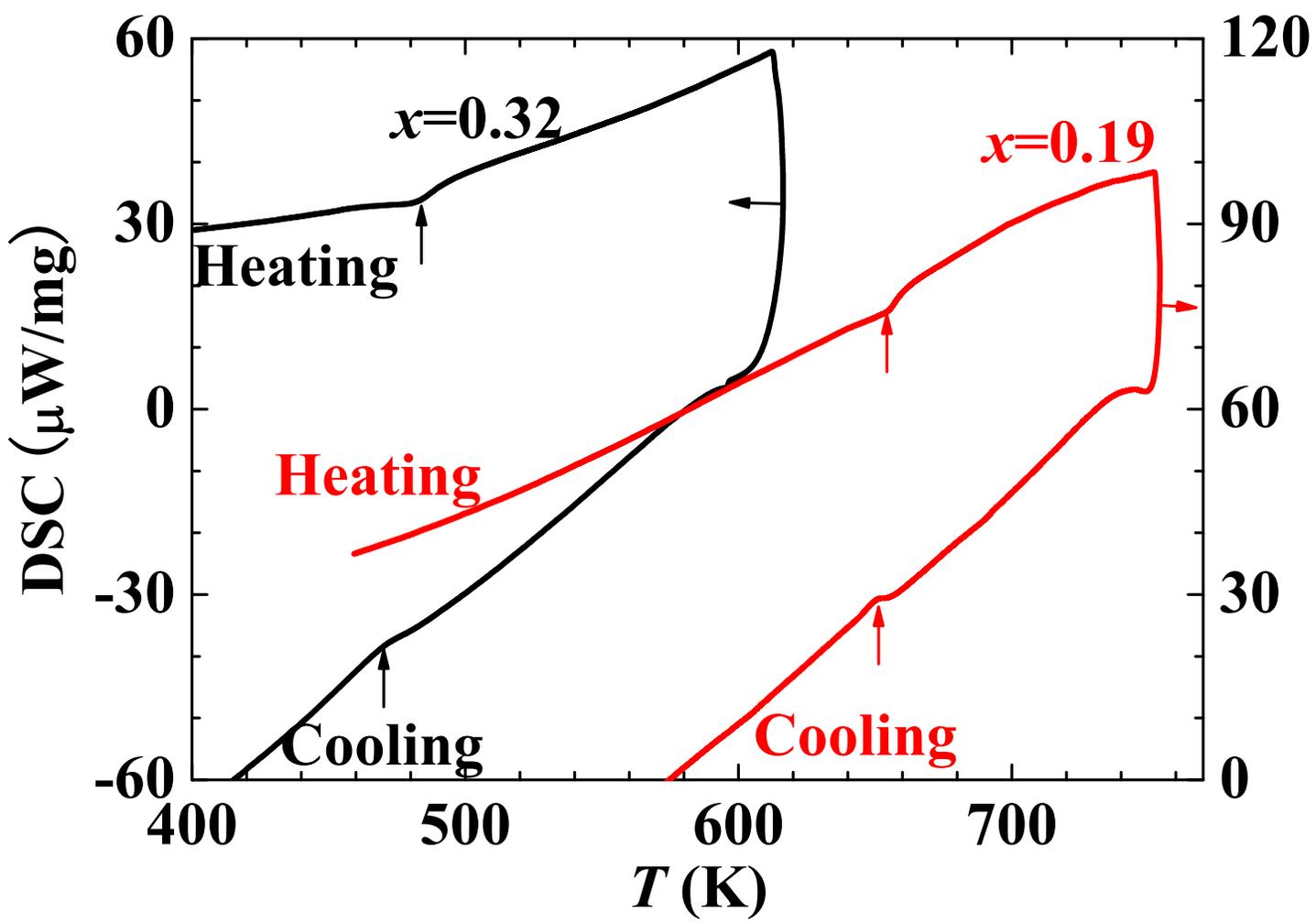
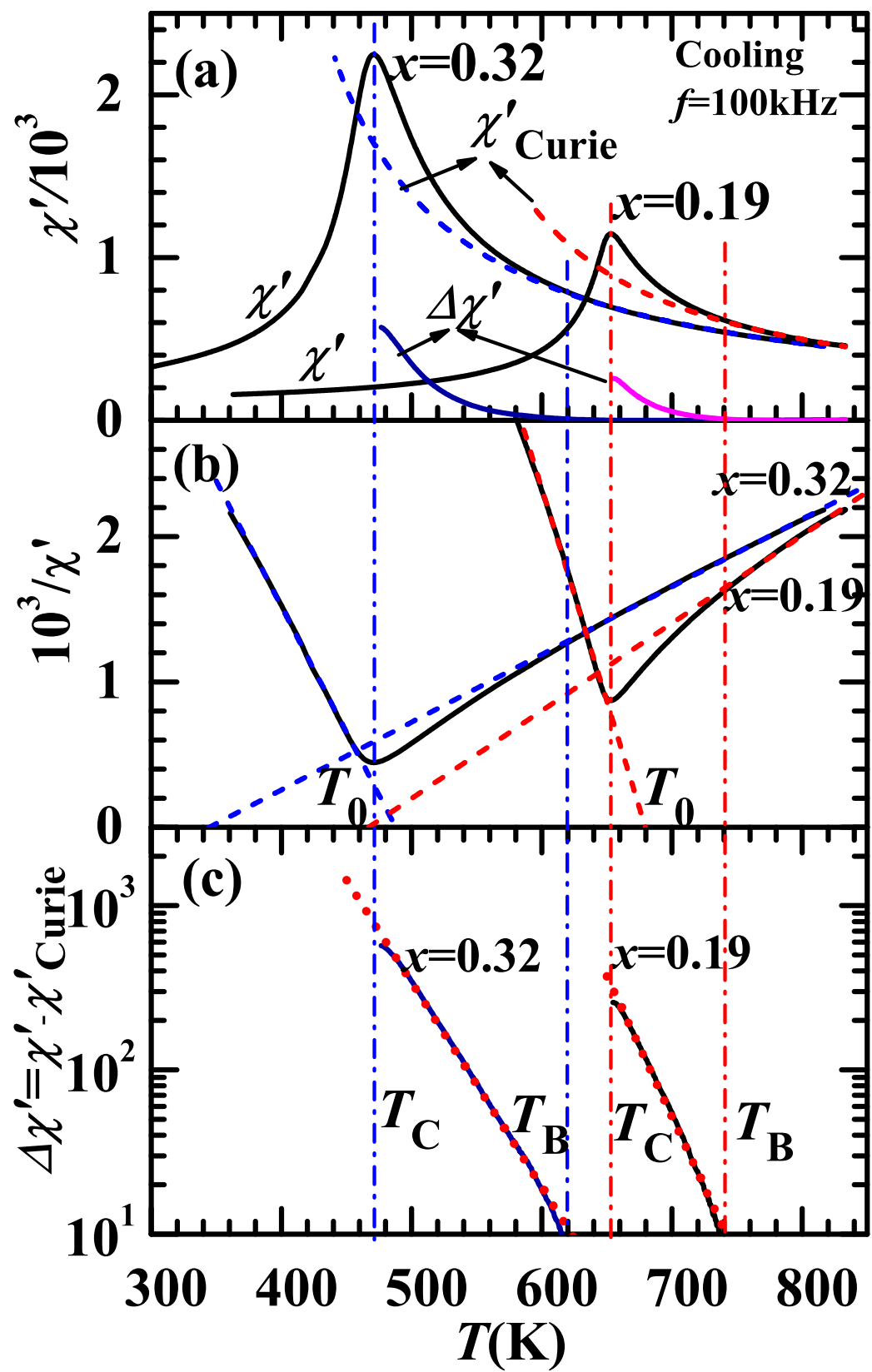
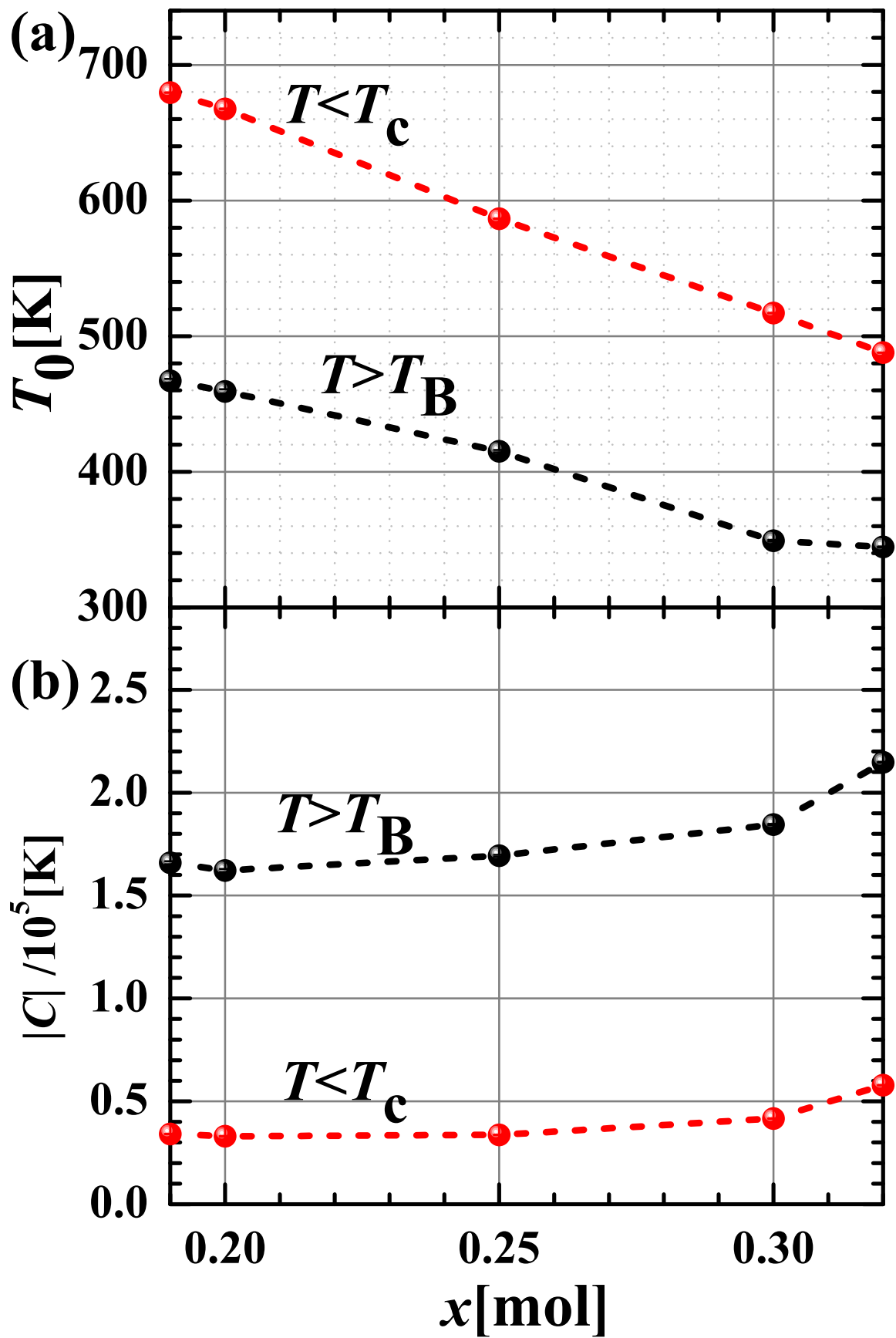
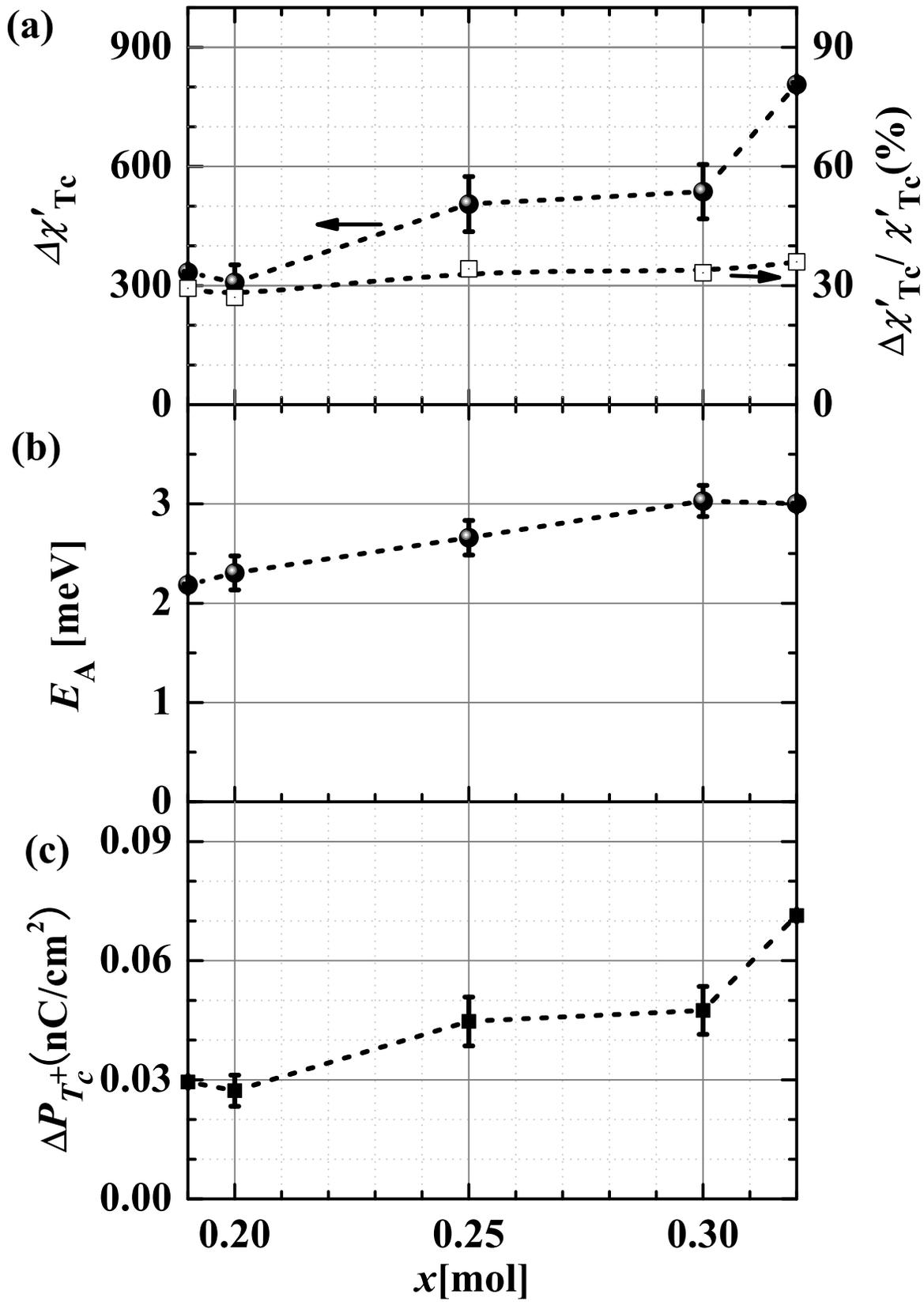
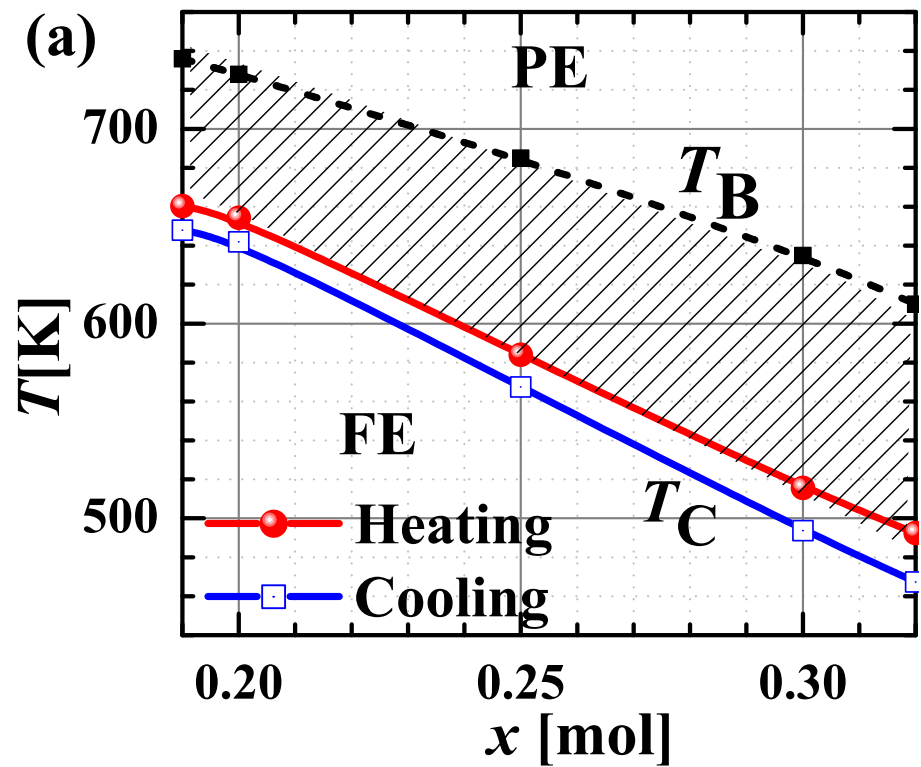
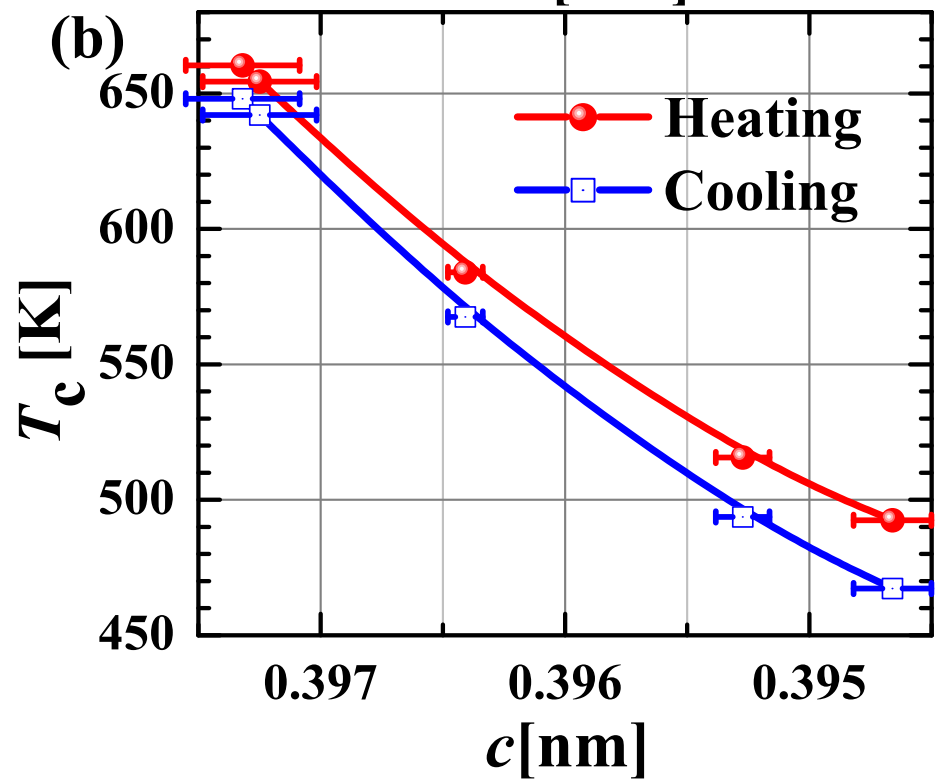
\end{document}